\documentclass[12pt]{iopart}
\usepackage{iopams}
\usepackage{epsfig}
\usepackage{textcomp}
\bibstyle{plain}
\begin{document}

\title[Emergence of multilevel selection in the prisoner's dilemma game on coevolving...]{Emergence of multilevel selection in the prisoner's dilemma game on coevolving random networks}

\author{Attila Szolnoki$^1$ and Matja{\v z} Perc$^2$}
\address{$^1$Research Institute for Technical Physics and Materials Science, P.O. Box 49, H-1525 Budapest, Hungary\\ $^2$Department of Physics, Faculty of Natural Sciences and Mathematics, University of Maribor, Koro{\v s}ka cesta 160, SI-2000 Maribor, Slovenia}
\ead{szolnoki@mfa.kfki.hu, matjaz.perc@uni-mb.si}

\begin{abstract}
We study the evolution of cooperation in the prisoner's dilemma game, whereby a coevolutionary rule is introduced that molds the random topology of the interaction network in two ways. First, existing links are deleted whenever a player adopts a new strategy or its degree exceeds a threshold value, and second, new links are added randomly after a given number of game iterations. These coevolutionary processes correspond to the generic formation of new and deletion of existing links that, especially in human societies, appear frequently as a consequence of ongoing socialization, change of lifestyle or death. Due to the counteraction of deletions and additions of links the initial heterogeneity of the interaction network is qualitatively preserved, and thus cannot be held responsible for the observed promotion of cooperation. Indeed, the coevolutionary rule evokes the spontaneous emergence of a powerful multilevel selection mechanism, which despite of the sustained random topology of the evolving network, maintains cooperation across the whole span of defection temptation values.
\end{abstract}

\maketitle

\section{Introduction}

Social dilemmas emerge if actions warranting individual success harm collective wellbeing \cite{macypnas02}. While cooperative behavior is universally regarded as the rational strategy leading away from the impending social decline, its evolution in groups of selfish individuals is puzzling \cite{axelrod84}. Evolutionary game theory is frequently used as the framework within which answers to the puzzle are sought \cite{maynardsmith82}, and the prisoner's dilemma game seems particularly suited in this respect. In this game mutual cooperation yields the highest collective payoff that is equally shared between the players. However, defectors will do better as individuals if their opponents cooperate. In well mixed populations selfishness always leads to mutual defection \cite{hofbauer88}, whereby players remain empty-handed and the society suffers. The pivotal study that launched a spree of activity aimed towards resolving the dilemma is due to Nowak and May \cite{nowaknat92}, who showed that spatial structure may maintain cooperative behavior in the prisoner's dilemma game. Further mechanisms promoting cooperation are kin selection \cite{hamiltonjtb64}, direct and indirect reciprocity \cite{axelrodsci81, fehrnat02}, as well as group \cite{maynardsmithn64, wilsonan77, wilsonare83} and multilevel selection \cite{traulsenpnas06, traulsenbmb08}, as recently reviewed in \cite{nowaksci06}. Related specifically to the present work is the promotion of cooperation via multilevel selection, which can be related to group selection \cite{wilsonbbs98}, although the latter term is frequently addressed in a rather problematic fashion (see \cite{westjeb08} for a recent recap).

Particularly vibrant in recent years has been the subject of evolutionary games on complex networks \cite{abramsonpre01, szabopr07}. High cooperation levels reported on scale-free networks highlighted the beneficial impact of heterogeneity that characterizes their degree distribution \cite{santosprl05}. Notably, the promotive impact of heterogeneous states on the evolution of cooperation has been reported also in other contexts \cite{percpre08, santosnat08}. Although several studies have since elaborated on different aspects of strategy adoption on complex networks \cite{ohtsukiprl07, masudaprsb07, puschpre08, assenzapre08, rongpre08, kupermanepjb08, wupre09, dupa09, fupre09, yangnjp09}, open questions remain. Foremost, it is still of interest to investigate how the coevolution of networks affects the evolution of strategies. Following the earlier works in the context of evolutionary game theory \cite{ebelpre02, zimmermannpre04, pachecoprl06}, it has recently been shown that highly heterogeneous interaction networks may evolve spontaneously from simple coevolutionary rules \cite{santosplos06, lipre07, poncelaplos08, szolnokiepl08, poncelanjp09}, and moreover, processes like appropriate reactions to adverse ties \cite{segbroeckbmc08, segbroeckprl09}, reputation-based partner choice \cite{fupre08}, as well as increase of teaching activity \cite{szolnokinjp08, szolnokiepjb09}, have all been considered as coevolutionary rules that can promote cooperative behavior. This subject is intimately connected with the seminal works on network growth \cite{barabasisci99, albertrmp02} and their resilience to error and attack \cite{albertnat00, cohenprl00, callawayprl00, cohenprl01} as means to alter the topology in order to affect the spread of epidemics and viral infections \cite{pastorpre02, zanettepa02, barthelemyprl04} in an efficient way \cite{cohenprl03}, as recently argued in \cite{percnjp09}. Moreover, the evolution \textit{on} networks is increasingly often accompanied also by the evolution \textit{of} networks not just in the context of evolutionary games \cite{holmeprl06, grossprl06}, but indeed networks are to be seen as evolving entities that may substantially influence all dynamical processes that are taking place on them \cite{dommarepjst08, grossjrsi08}.

Aim of the present work is to show that coevolutionary rules molding the interaction network can influence the evolution of cooperation not via the emergence of strong degree heterogeneity, as reported earlier \cite{poncelaplos08, szolnokiepl08, poncelanjp09}, but may indeed introduce processes that affect the macroscopic dynamics of strategy adoption. We introduce a simple strategy-independent coevolutionary rule that qualitatively preserves the Poissonian degree distribution of the initial random interaction network. The coevolutionary rule involves the deletion of existing links upon the adoption of a new strategy or exceeding of a given maximal degree, and the addition of new links after every $\tau$ full Monte Carlo steps. These coevolutionary additions and deletions of links are motivated by the fact that, especially in human but also in animal societies, ties between members of the population change in time. In particular, the formation of links is common~--~\textit{e.g.} we socialize, we make friends~--~thus, in the course of a lifetime we certainly form many new ties with others. We take this into account in our model by adding randomly a new link to each player every $\tau$ full Monte Carlo steps. Note that since we add new links every $\tau$ game iterations, the number of links an individual accumulates over time can be directly linked with age. This is why extinction (death) of players is considered by deleting existing links whenever the degree of a player exceeds a threshold value. And finally, there are situations in life when an individual changes a significant part of its existence. Within our model we consider this to be the player's strategy, while in reality examples that can be considered as related are changing the lifestyle, moral values or political orientation, all of which typically lead to restructuring of one's connections with others. In accordance, we therefore delete existing links of the player that changes its strategy. From a biological viewpoint, the latter act can be linked with an invasion of the subordinate species and the subsequent replacement by a newborn of the victor.

The proposed coevolutionary rule thus generically describes the formation and deletion of links with other members of the society in a general and strategy-independent (the same rules apply for cooperators and defectors) manner. Jointly, the two processes of link addition and deletion annihilate each other's fingerprint on the heterogeneity of the degree distribution, thereby eliminating network structure as a potentially decisive factor by the evolution of cooperation. Remarkably though, we demonstrate that the coevolutionary rule spontaneously evokes a dynamical mechanism that can be interpreted as a multilevel selection (see \textit{e.g.} \cite{traulsenpnas06}) in that, on the macroscopic level, groups of cooperators have much better chances of dissemination than groups of defectors, whereas on the microscopic level, defectors within a given group are superior to cooperators. The spontaneous emergence of multilevel selection within the proposed coevolutionary model leads to complete dominance of cooperators across the whole span of the temptation to defect provided $\tau$ is large enough. The study thus supplements previous works examining the impact of different coevolutionary rules, and moreover, demonstrates that heterogeneity can be of secondary importance by the evolution of cooperation on complex networks.

The remainder of this paper is organized as follows. In the next section we describe the employed evolutionary prisoner's dilemma game and the protocol for the coevolution of the random network. Section 3 is devoted to the presentation of the main findings, whereas in the last section we summarize conclusions based on them.

\section{Mathematical model}

Here the prisoner's dilemma game is used as a representative example of a social dilemma, whereby we adopt the same parametrization as proposed in \cite{nowaknat92}. Accordingly, the game is characterized by the temptation to defect $T=b$, reward for mutual cooperation $R = 1$, and punishment $P$ as well as the suckers payoff $S$ equaling $0$, whereby $1 < b \leq 2$. In the game two cooperators facing one another acquire $R$, two defectors get $P$, whereas a cooperator receives $S$ if facing a defector who then gains $T$. This can be summarized succinctly by the corresponding payoff matrix given in Table~\ref{pmat}.
\begin{table}
\caption{Payoff matrix of the prisoner's dilemma game. Strategies in rows get the depicted payoff when playing the game with the strategies in columns.}
\centering
\vspace{0.5cm}
\begin{tabular}{lll}
& \vline \,\, $C$ & \,\,\,$D$ \\
\hline
$C$ & \vline \,\,\,$1$ &\,\, $0$ \\
$D$ & \vline \,\,\,$b$ &\,\, $0$ \\
\end{tabular}
\label{pmat}
\end{table}
It is worth noting that the prisoner's dilemma we use is not strict in that $P$ is not strictly larger than $S$, but the qualitative dynamics of the game is thereby not affected. Initially, each player $x$ is designated either as a cooperator $(s_x=C)$ or defector $(s_x=D)$ with equal probability, and is placed on a random network that is constructed from $N$ individuals so that the average and minimal degree are $k_{avg}=4$ and $k_{min}=1$ (no player is detached), respectively. Duplicate links are also omitted. Evolution of the two strategies is performed in accordance with the Monte Carlo simulation procedure comprising the following elementary steps. First, a randomly selected player $x$ acquires its payoff $p_x$ by playing the game with all its $k_x$ neighbors. Next, one randomly chosen neighbor of $x$, denoted by $y$, also acquires its payoff $p_y$ by playing the game with all its $k_y$ neighbors. Last, if $p_x > p_y$ player $x$ tries to enforce its strategy $s_x$ on player $y$ in accordance with the probability $W(s_x \rightarrow s_y)=(p_x-p_y)/b k_q$, where $k_q$ is the largest of the two degrees $k_x$ and $k_y$. This proportional imitation rule \cite{helbing98} is used frequently when players on heterogeneous networks have different degrees \cite{santosprl05, assenzapre08, rongpre08, wupa07}. Notably, an alternative to the latter is the Fermi strategy adoption rule \cite{szabopre98}, which enables tuning of the selection intensity via a single parameter $K$ towards the weak selection limit \cite{wildjtb07, altrocknjp09}. However, caution should be exercised when applying it on heterogeneous networks since due to the different degrees of participating players the temperature $K$, and thus also the selection intensity, effectively varies from one strategy adoption to the other. We therefore use the simplified strategy adoption rule based on proportional imitation, but note that similar results as will be reported below are expected also for reasonably weak selection. However, if the strategy adoption process becomes comparable to the flip of a coin the below reported multilevel selection may be impaired. In accordance with the applied Monte Carlo procedure incorporating the random sequential update, each player is selected once on average during a full simulation step (MCS).

\begin{figure}
\begin{center} \includegraphics[width = 15.5cm]{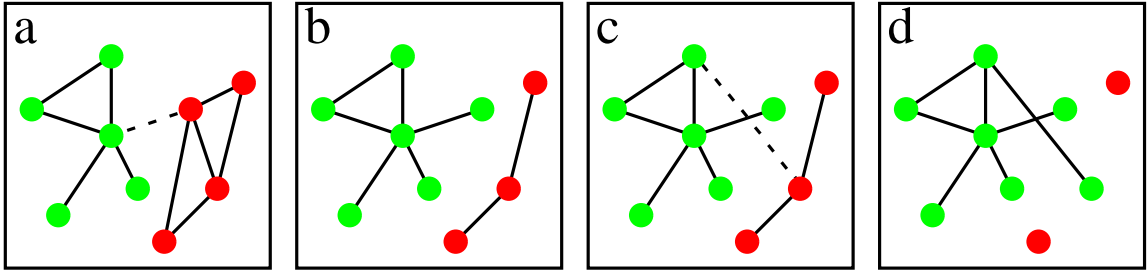}
\caption{\label{fig_loose} The central green player in panel (a) passes strategy to the red player that is connected dashed. In accordance with the proposed coevolutionary rule, in panel (b) the invaded player (now green) looses all its links, keeping only the one with the donor of the new strategy. Panels (c) and (d) depict the same process again for a different pair, whereby it is important to note that the dashed links may be a consequence of the random additions entailed in the model, occurring every $\tau$ full Monte Carlo steps. We will indeed show that the coevolutionary rule leads to the isolation of cooperating and defecting groups of players [as depicted in panels (a) and (c) if not taking into account the dashed links], and that the appropriately tuned frequency of adding new links (determined by $\tau$) triggers multilevel selection based on which groups of cooperators (green) prevail against groups of defector (red). The complete sequence of panels from (a) to (d) depicts this process schematically, and we will refer back to it when presenting the results.}
\end{center}
\end{figure}

\begin{figure}
\begin{center} \includegraphics[width = 8.5cm]{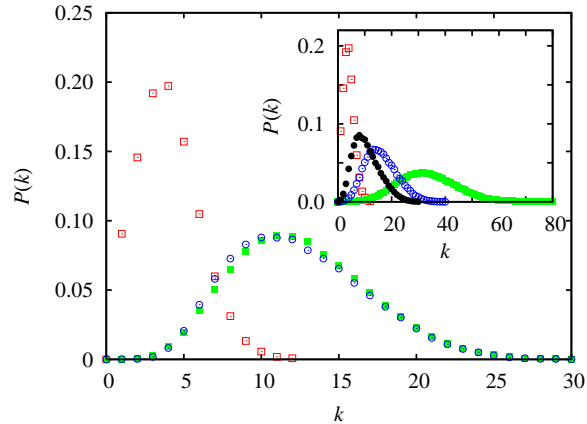}
\caption{\label{fig_dist} Degree distributions of interaction networks for different values of $b$ and $\tau$. The main panel features results obtained at $\tau=70$ for $b=1.6$ (filled green squares) and $b=2.7$ (open blue circles). Open red squares depict the initial degree distribution of the random interaction network (the same in the inset). Although the Poissonian outlay is preserved, the interaction network may warrant full cooperator ($b=1.6$; filled green squares) or full defector ($b=2.7$; open blue circles) dominance (see also Fig.~\ref{fig_psi} below), thus rendering the heterogeneity of the interaction network irrelevant for understanding the outcome of the main evolutionary process. The inset demonstrates the qualitative prevalence of the Poissonian degree distribution in dependence on $\tau$ ($30$~--~filled green squares; $70$~--~open blue circles; $500$~--~filled black circles) at $b=2$. The employed system size was $N=10^4$ and the values of $P(k)$ were obtained as averages over 100 independent realizations.}
\end{center}
\end{figure}

In addition to the evolution of the two strategies, a coevolutionary rule is implemented as follows. Whenever player $x$ adopts a new strategy all its links, except from the one with the donor of the new strategy, are deleted. Hence, in addition to adopting a new strategy, the player is separated from its former allies and gets $k_x=1$. This process of strategy adoption and simultaneous link deletion is demonstrated in Fig.~\ref{fig_loose}, and can be motivated by the fact that changes in lifestyle, moral values, political orientation or religious beliefs frequently result in deletion of existing ties we have formed with others. To counteract the depletion of links that constitute the random network, all individuals are allowed to form a new link with a random player with which they are not yet connected. The latter process, happening after every $\tau$ full Monte Carlo steps, corresponds to the continuous process of socialization or making of new friends, which typically entails the formation of new links. We also take into account aging, and accordingly, as soon as $k_x$ reaches a threshold $k_{max}$, player $x$ dies and is replaced by a newborn having the same strategy and keeping a single randomly selected link from its predecessor to assure connectedness. Note that since new links are added every $\tau$ game iterations, the number of links a player accumulates over time can be directly linked with age. Within the current work $k_{max}$ does not play a decisive role and was simply chosen large enough so as not to influence the initial random network topology and the subsequent evolution of cooperation. The presented results were obtained on networks hosting $N=10^4$ players, for which $k_{max}=500$ has proven to be sufficiently large. Importantly, due to the continuous additions and deletions of links the initial Poissonian outlay of the degree distribution is qualitatively preserved irrespective of $b$ and $\tau$, as shown in Fig.~\ref{fig_dist}. It is intriguing that, although the coevolutionary rule does not affect the heterogeneity of the interaction network in a significant manner, the promotion of cooperation depends crucially on $\tau$, as can be inferred from the caption of Fig.~\ref{fig_dist}. These motivational results presented in Fig~\ref{fig_dist} will be explained in the next section. It is also worth mentioning that the process illustrated in Fig.~\ref{fig_loose} may occasionally result in detached players that originally formed the neighborhood of the invaded player, as shown in Fig.~\ref{fig_loose}(d). In this case we relinked the detached player randomly onto the network. Next, we will systematically analyze the evolution of cooperation in dependence on $b$ and $\tau$, which are the two crucial parameters within the proposed coevolutionary model. Notably, the evolution of cooperation on static random networks has been studied in \cite{duranpd05}, and the reader is refereed there for comparisons with respect to the above proposed coevolutionary model.

\section{Results}

\begin{figure}
\begin{center} \includegraphics[width = 8.5cm]{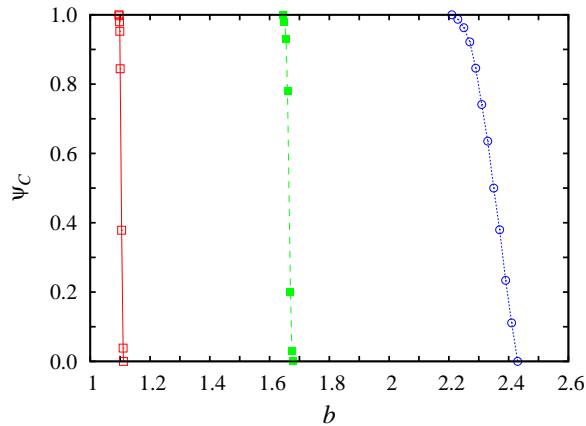}
\caption{\label{fig_psi} $\psi_C$ in dependence on $b$ for $\tau=1$ (open red squares), $\tau=10$ (filled green squares) and $\tau=70$ (open blue circles). Cooperation is promoted substantially as $\tau$ increases, resulting in full dominance of cooperators up to $b=2.2$ (beyond the prisoner's dilemma range) at $\tau=70$. Note that for each $\tau$ the fixation probability of cooperators is always one before it breaks down (see also main text for details). The employed system size was $N=10^4$, and the absorbing state was reached within $2 \cdot 10^2 - 5 \cdot 10^4$ full Monte Carlo steps depending on $b$ and $\tau$. Final values of $\psi_C$ were obtained as averages over $10^3$ independent realizations. Lines are just to guide the eye.}
\end{center}
\end{figure}

Within the coevolutionary model the final outcome of the prisoner's dilemma game is always an absorbing $C$ or $D$ state. We note that this is an inherent property of the coevolutionary model that prevails irrespective of the system size $N$. In order to account for the resulting fluctuating output near sharp transition points, we depict in Fig.~\ref{fig_psi} not the density of cooperators $\rho_C$, but rather the probability $\psi_C$ that the final state is $\rho_C=1$, whereby the latter is determined within $1000$ independent runs for each particular combination of $b$ and $\tau$. Foremost, Fig.~\ref{fig_psi} depicts the effect of different values of $\tau$ on the evolution of cooperation. Clearly, the coevolutionary process promotes cooperation extremely effectively, resulting in full dominance of cooperators up to $b=1.1$ at $\tau=1$, $b=1.6$ at $\tau=10$ and up to $b=2.2$ at $\tau=70$, where the latter value of the temptation to defect is already past the span of the prisoner's dilemma game as it is considered here. Given that the continuous deletions and additions of links, entailed in the coevolutionary process, qualitatively preserve the initial random topology and the heterogeneity of the interaction networks (see Fig.~\ref{fig_dist}), the observed promotion of cooperation cannot be attributed to mechanisms that rely on heterogeneous environments reported earlier \cite{santosprl05, percpre08}. Thus, instead we have to look for additional clues that may explain the doom of defectors by large values of $\tau$, as evidenced in Fig.~\ref{fig_psi}.

\begin{figure}
\begin{center} \includegraphics[width = 8.5cm]{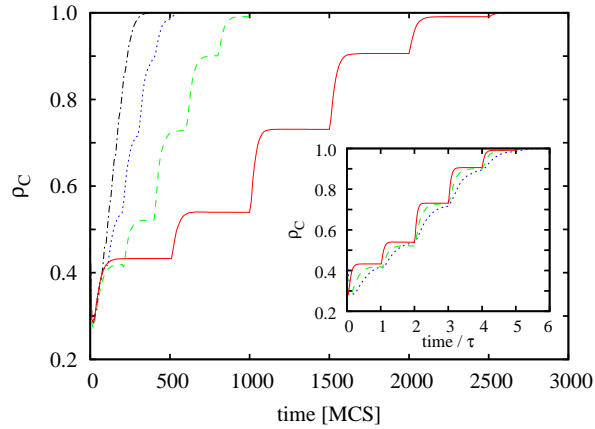}
\caption{\label{fig_time} Temporal evolution of $\rho_C$ at $b=1.5$ for $\tau=30$ (dash-dotted black line), $\tau=100$ (dotted blue line), $\tau=200$ (dashed green line) and $\tau=500$ (solid red line). Steps in the temporal evolution (time intervals during which $\rho_C$ is constant), which can be observed by larger values of $\tau$, indicate the emergence of multilevel selection in the examined model. For easier comparisons of the duration of dormant phases the inset features the time axes scaled with $\tau$ pertaining to each of the depicted curves (note that $\tau=30$ is not shown for clarity). Larger values of $\tau$ clearly evoke longer time intervals during which $\rho_C$ is constant. The employed system size was $N=10^4$.}
\end{center}
\end{figure}

Striving towards a mechanism that could explain the promotion of cooperation, we show in Fig.~\ref{fig_time} temporal courses of $\rho_C$ at $b=1.5$ for increasing values of $\tau$ from left to right. First, it is worth noticing that indeed the absorbing cooperative state is reached fairly quickly, \textit{i.e.} within a few thousand full Monte Carlo steps. But most importantly in Fig.~\ref{fig_time}, we point out the emergence of time intervals during which $\rho_C$ is constant. This cascade-like feature becomes increasingly pronounced as $\tau$ increases. Namely, at $\tau=30$ (dash-dotted black line) it is practically absent, whereas at $\tau=500$ (solid red line) the cumulative duration of dormancy of $\rho_C$ surpasses that of active phases. We argue that the reason for the emergence of these inactive windows lies in the newly introduced coevolutionary process, which dictates continuous deletion of links that occurs always when a player adopts a new strategy, and also if its degree reaches $k_{max}$. If $\tau$ is sufficiently large, meaning that new links are added rarely, the deletions of links lead to the emergence of homogeneous and virtually isolated groups of players [as schematically depicted in Fig.~\ref{fig_loose}(a) and ~\ref{fig_loose}(c) not taking into account the dashed links]. These groups remain inactive for as long as it takes for the newly added links [dashed link in Fig.~\ref{fig_loose}(a) and ~\ref{fig_loose}(c)] to reconnect them with one another, of which duration is roughly equivalent to $\tau$ (see the inset of Fig.~\ref{fig_time}). It is important to note that during the inactive phase there are practically no strategy transfers taking place, and thus the main source of link deletions is disabled. Consequently, the addition of new links can gradually reconnect the detached groups, which then again triggers an avalanche of strategy adoptions [see Figs.~\ref{fig_loose}(b) and ~\ref{fig_loose}(d)], which in turn starts the whole process anew, until eventually an absorbing state is reached. Thus, the temporal plots in Fig.~\ref{fig_time} arguably evidence the spontaneous emergence of multilevel selection, similarly as proposed recently in \cite{traulsenpnas06}, due to the introduction of the proposed coevolutionary rule. We emphasize that the groups are not introduced a priori but eventually emerge spontaneously for appropriate values of $\tau$. During the dormant phases isolated groups of cooperators can enhance their strength, while groups of defectors weaken as there is nobody to exploit. Notably, from the viewpoint of vulnerability there is no difference between defectors having a small or large degree. Thus, as soon as the two types of groups reestablish a sufficiently strong interconnectedness, cooperators can successfully invade the defectors, thereby gradually increasing  the cooperative domains. This is schematically depicted in Fig.~\ref{fig_loose} if considering the cooperators to be green and defectors to be red. The latter processes manifest as rather steep jumps in the temporal traces of $\rho_C$ by large enough $\tau$, which are then again followed by dormancy since the many strategy adoptions anew lead to isolation of homogeneous groups of players. At small values of $\tau$ (dash-dotted black line in Fig.~\ref{fig_time}), however, these steps are absent, which indicates that the addition of new links is too fast for the homogeneous groups to become isolated enough to trigger the multilevel selection. Importantly, a thorough isolation is necessary (compare the curves in the inset of Fig.~\ref{fig_time}), since without it groups hosting both strategies are susceptible to an overrule by defectors. Namely, in the small $\tau$ region the fast additions and deletions of links yield mean-field type conditions that are arguably harmful for cooperators.

\begin{figure}
\begin{center} \includegraphics[width = 8.5cm]{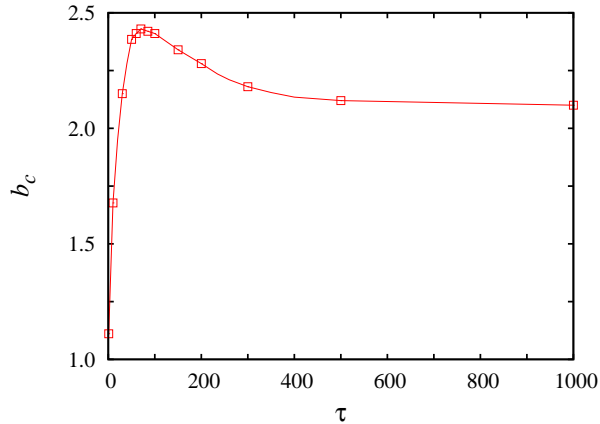}
\caption{\label{fig_bc} Critical $b_c$, bordering $D$ phase, in dependence on $\tau$. The best promotion of cooperation is warranted approximately at $\tau=70$. The employed system size was $N=10^4$ and the values of $b_c$ were obtained as averages over $10^3$ independent realizations. Line is just to guide the eye.}
\end{center}
\end{figure}

The above interpretation can be corroborated nicely by results presented in Fig.~\ref{fig_bc}, where the critical temptation to defect $b_c$ is plotted in dependence on $\tau$. We determined $b_c$ as the lowest value of $b$ where an absorbing $D$ state is attained with probability one. It can be observed that, as argued above, by small values of $\tau$, where multilevel selection cannot be fully established, the promotion of cooperation is indeed marginal. However, it improves steeply and reaches a maximum at approximately $\tau=70$. According to the temporal outlays presented in Fig.~\ref{fig_time}, this is roughly the value of $\tau$ where the multilevel selection becomes fully pronounced. In particular, note that the steps by $\tau=100$, and even by $\tau=30$, are practically just as frequent as by $\tau=500$. Crucially, however, the dormant phases are shorter by smaller $\tau$. Thus, as $\tau$ increases past the optimal value, solely the dormant phases are prolonged, yet the multilevel selection remains equally intense. Therefore, by values of $\tau$ past the optimum the defectors have more time to overtake individual groups during the dormant phases. Note that the isolation of homogeneous groups from the opposite strategy is never complete since neither individual players nor groups can become fully detached from the network. The prolongation of dormant states thus results in a slight decrease and subsequent saturation of $b_c$. Nevertheless, due to the remaining intact multilevel selection even at large $\tau$, the promotion of cooperation is still significant, letting the critical temptation to defect hoover comfortably over the maximal $b$ permissible within the prisoner's dilemma game.

\section{Summary}

In sum, we have elaborated on the evolution of cooperation in the prisoner's dilemma game on evolving random networks. Foremost, we have shown that a simple coevolutionary rule, qualitatively preserving the initial heterogeneity of the interaction network, may spontaneously evoke a powerful multilevel selection mechanism that promotes cooperation across the whole span of values of temptation. The extend of the promotive effect depends on a single parameter $\tau$, determining the frequency of forming new links between randomly chosen but not yet linked players. While this dependence exhibits a saturating outlay with a local maximum at $\tau=70$, the cooperation promoting mechanism remains intact also for substantially larger $\tau$ across the whole span of $b$. The optimum emerges as a result of the fully developed multilevel selection, which is not attainable for small values of $\tau$ because too frequent additions of new links hinder the formation of isolated homogeneous groups and yield mean-field type conditions. On the other hand, as additions of new links become rare ($\tau$ is increased), the saturating effect of cooperation promotion sets in due to the emergence of prolonged delays between periods of active multilevel selection, during which defectors may be able to invade and overtake individual cooperative groups. Our work demonstrates that simple strategy independent coevolutionary rules may spontaneously evoke dynamical mechanisms that affect the adoption of strategies on the macroscopic level of evolutionary game dynamics. This presently manifests as multilevel selection that strongly promotes cooperation in the prisoner's dilemma game.

\ack
This work was supported by the Slovenian Research Agency (grant Z1-2032-2547), the Hungarian National Research Fund (grant K-73449), the Bolyai Research Scholarship, and the Slovene-Hungarian bilateral incentive (grant BI-HU/09-10-001). Discussions with Gy{\"o}rgy Szab{\'o} are gratefully acknowledged.

\end{document}